\documentstyle[11pt,newpasp,twoside,epsf]{article}
\reversemarginpar

\begin{document}
\title{A search for high redshift clusters associated with radio
galaxies at 2 $<$ z $<$ 4}

\author{J.D. Kurk, B.P. Venemans, H.J.A. R\"ottgering, G.K. Miley}
\affil{Sterrewacht Leiden, P.O. Box 9513, 2300 RA, Leiden, The Netherlands}
\author{L. Pentericci}
\affil{Max-Planck-Institut f\"ur Astronomie, K\"onigstuhl 17, D-69117, 
Heidelberg, Germany}

\begin{abstract}
High redshift radio galaxies are amongst the most massive galaxies in
the early Universe and have properties expected from central galaxies
in forming clusters. We are carrying out an observational programme on
the VLT to find and study galaxy proto clusters around radio galaxies
at redshifts $2 < z < 4$. First, we use narrow band imaging to select
candidate galaxies which show excess Ly$\alpha$ emission at redshifts
similar to the central radio galaxy. Then, we use multi object
spectroscopy to confirm the redshifts of these candidates and measure
the velocity dispersion of the cluster members. Our goal is to observe
a sample of $\sim 10$ targets and investigate galaxy overdensities as
a function of redshift. Here, we report on the current progress of the
programme and show some preliminary results which include the
discovery of a structure of galaxies at redshift 4.1.
\end{abstract}

\section{Introduction}

Searching for clusters of galaxies or their progenitors (proto
clusters) at redshifts $> 2$ is important for constraining galaxy
evolution, large scale structure formation and cosmological models
(e.g.\ Giavalisco et al.\ 1998, Bahcall, Fan, \& Cen 1997). Radio
galaxies can act as beacons of these distant structures.

Luminous radio galaxies are the oldest and most massive known galaxies
in the early Universe (De Breuck et al.\ 2000). In standard
cosmological models massive galaxies form in the deepest of the first
potential wells, in which - at either a later or earlier time
depending on the galaxy formation model - other galaxies will form to
assemble a cluster. There exists a diverse collection of evidence that
high redshift radio galaxies (HzRGs) are actually residing in dense
cluster environments, which includes: (i) the excess of companion
galaxies around HzRGs (e.g.\ R\"ottgering et al.\ 1996), (ii) the
extreme radio rotation measures of HzRGs indicating that they are
embedded in hot magnetized gas (Carilli et al.\ 1997), and (iii) giant
Ly$\alpha$\ emitting gas halos surrounding HzRGs which can extent up
to 200 kpc (e.g.\ Kurk et al.\ 2001).

In principle, using narrow band imaging in rest-frame Ly$\alpha$ star
forming galaxies can be found effectively over a narrow redshift
range. However, since there is only a small amount of dust needed to
extinguish Ly$\alpha$\ emission from recombining hydrogen ions, it was
not clear whether Ly$\alpha$ photons could escape at all from star
forming galaxies (Pritchet 1994). Now, using large telescopes several
searches for Ly$\alpha$\ emitters are underway and show good
results (Rhoads et al.\ 2000, Stiavelli et al.\ 2001). With the
presumption that powerful radio galaxies are located in overdense
regions, we have started a search for high redshift clusters using the
narrow band imaging technique to target Ly$\alpha$ emitting galaxies
around radio galaxies at redshifts $2 < z < 4$.

\section{The sample}

From our collection of almost 150 radio galaxies at $z > 2$, we have
chosen eight objects with redshifts suitable for Ly$\alpha$\ imaging
with the available narrow band VLT/FORS filters: four at $z \sim 2$
and three at $z \sim 3$, while for two more galaxies at $z = 4.1$ a
custom filter was manufactured. These galaxies were also selected on
their large radio, optical/IR continuum and Ly$\alpha$\ luminosities,
which are characteristics of massive galaxies. For each of the ten
radio galaxy fields we (plan to) do spectroscopy of $\sim 40$ emitter
candidates and aim to confirm $> 15$ emitters. With this sample, we
can study what fraction of HzRGs are located in galaxy overdensities
and how properties of these overdensities (e.g.\ velocity dispersions,
sizes) and its cluster-galaxies (e.g.\ rotation curves, star formation
rates) change with redshift. Simultaneously, we can study the giant
gas halos associated with the radio galaxies.

Currently, imaging of five fields has been carried out, resulting in
the detection of fifteen to sixty emitters in each field with
Ly$\alpha$ rest frame equivalent width $> 20$ \AA\ and significant
signal-to-noise. Multi object spectroscopy has been completed for two
fields: MRC 1138-262 and TN J1338-1942. The latter has the largest
redshift of our sample and is described in the following section. The
former was the subject of our pilot project and is described in detail
by Pentericci et al.\ (this volume). Briefly, the existence of 14
Ly$\alpha$ emitters in this field was confirmed, with redshifts close
to the radio galaxy but distributed in two groups (grey histogram in
Fig.\ 1).

\section{TN J1338-1942 at z = 4.11}

About 25 candidates with above mentioned selection criteria were
discovered in the field around TN J1338-1942. Subsequent spectroscopy
of 22 of these led to the confirmation of 20 Ly$\alpha$\ emitters at
$z \sim 4.1$ (Fig.\ 2). Their spatial distribution is inhomogeneous:
it is extended in the direction of the radio axis and Ly$\alpha$ halo
of the radio galaxy (Fig.\ 2). Their velocity distribution (black
histogram in Fig.\ 1) is narrower than that of MRC 1138-262 and has a
dispersion ($\sigma_{\rm v}$) of $\sim$ 350 km s$^{-1}$ which implies
a total virial mass (M$_{\rm vir}$) of $2 \times 10^{14}$ M$_{\sun}$,
assuming spherical symmetry with a virial radius (R) of 1.5 Mpc, using
M$_{\rm vir}$ = 5~R~$\sigma_{\rm v}^2$~/~G, where G is the gravitation
constant. To determine whether we actually observe an overdensity here
with respect to a field which does not contain a (proto) cluster, we
have compared the number density of emitters around 1338-1942 with
observations of blank fields by Cowie \& Hu (1998) and Rhoads et al.\
(2000). The number density of emitters around 1338-1942 is five times
higher than the field density and comparable to the density of
emitters around 1138-262.

\section{Conclusions and outlook}

\begin{figure}
\plotone{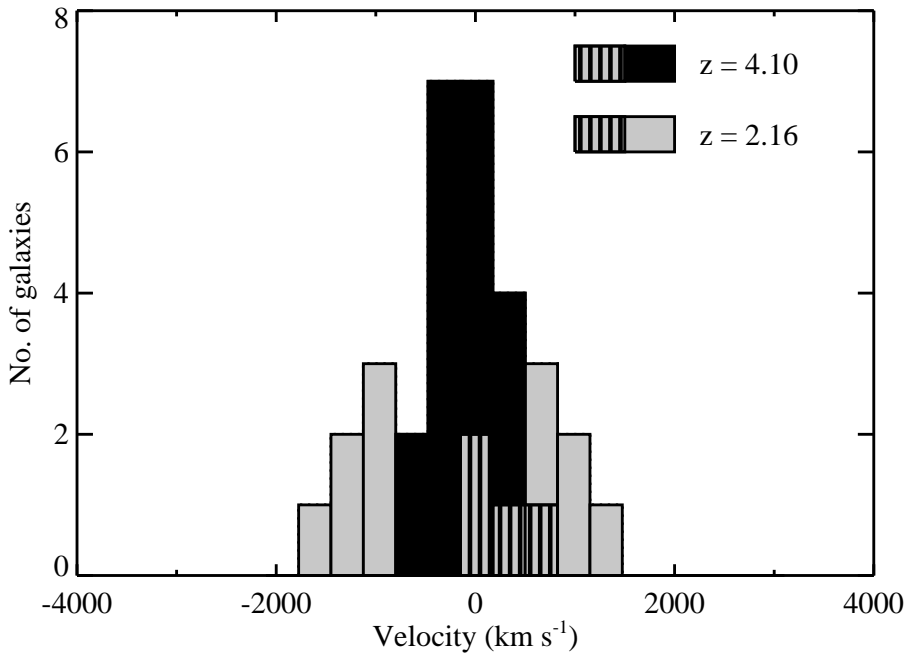}
\caption{Velocity distributions of confirmed emitters around MRC
1138-262 at $z=2.16$ (grey histogram) and TN J1338-1942 at $z=4.11$
(black histogram). The emitters around 1138 are clustered in two
groups, separated by $\sim 1800$ km s$^{-1}$, while the emitters
around 1338 have approximately a Gaussian distribution around the
radio galaxy.}
\end{figure}

Using narrow band imaging and subsequent multi object spectroscopy on
the VLT, we have confirmed the existence of concentrations of galaxies
around radio galaxies at $z=2.2$ and 4.1. These observations are
consistent with the idea that HzRGs are brightest cluster galaxies in
forming clusters. Furthermore, it shows that narrow band imaging at
the wavelength of redshifted Ly$\alpha$ is an effective method to find
(proto) clusters at high redshift. We expect to uncover several more
of these structures in the coming years.  Obviously, these systems
will be subject to further study at a multitude of wavelength bands,
contributing important clues to outstanding questions in galaxy
formation and cosmology.

\begin{figure}
\plottwo{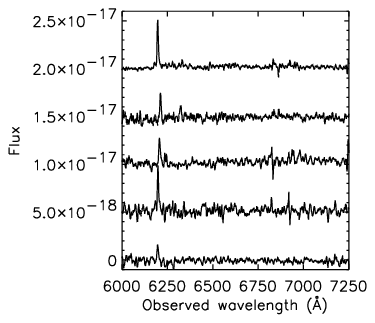}{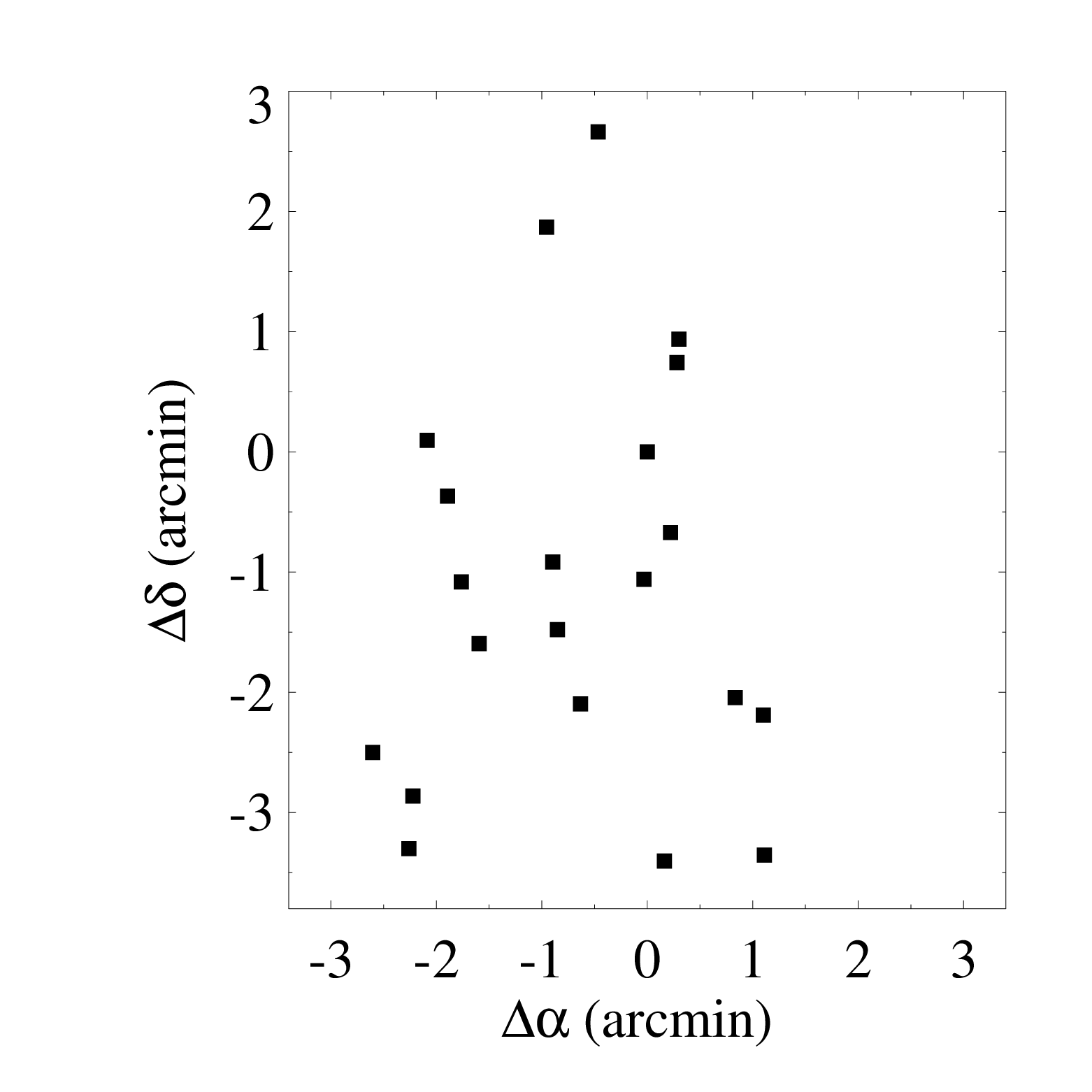}
\caption{\emph{Left} Spectra of five Ly$\alpha$ emitters found in the field of
TN J1338-1942. \emph{Right} Positions of the Ly$\alpha$ emitters around the
radio galaxy TN J1338-1942. The radio galaxy itself is located at the origin.}
\end{figure}

\section{References}
Bahcall,~N.~A., Fan,~X., Cen, R. 1997, ApJ, 485, L53\\
Carilli,~C.~L., R\"ottgering~H.~J.~A., van Ojik,~R., Miley,~G.~K., \& van
Breugel, W.~J.~M.\ 1997, ApJS, 109, 1\\
Cowie, L. L.\ \& Hu, E. M. 1998, AJ, 115, 1319\\
De Breuck,~C., van Breugel,~W., R\"ottgering,~H.~J.~A., \&
Miley,~G.~K.\ 2000, A\&AS, 143, 303\\
Giavalisco,~M., Steidel,~C.~C., Adelberger,~K.~L., Dickinson,~M.~E.,
Pettini,~M., \& Kellog,~M.\ 1998, ApJ, 503, 543 \\
R{\"o}ttgering,~H.~J.~A., West,~M.~J., Miley,~G.~K., \&
Chambers,~K.~C.\ 1996, A\&A, 307, 376\\
Kurk,~J.~D. et al. 2000, A\&A, 358, L1\\
Kurk,~J.~D., R\"ottgering,~H.~J.~A., Miley,~G.~K., \& Pentericci,~L.\
2001 in RevMexAA Conf.\ Proc.\ 'Emission Lines from Jet Flows', 
ed.\ W.\ Henney, W.\ Steffen, L.\ Binette, \& A.\ Raga, astro-ph/0102337\\
Pentericci, L. et al. 2000, A\&A, 361, L25\\
Pritchet, C. 1994, PASP, 106, 1052\\
Rhoads, J.~E., Malhotra, S., Dey, A., Stern, D., Spinrad, H., \&
Jannuzi, B.~T.\ 2000, ApJ, 545, L85\\ 
Stiavelli,~M., Scarlata,~C., Panagia,~N., Treu,~T., Bertin,~G., \&
Bertola,~F.\ 2001, ApJ in press, astro-ph/0105503 \\

\end{document}